# Violent Crime in London: An Investigation using Geographically Weighted Regression

Arman Sarjou

**Abstract**—Violent crime in London is an area of increasing interest following policing and community budget cuts in recent years. Understanding the locally-varying demographic factors that drive distribution of violent crime rate in London could be a means to more effective policy making for effective action. Using a visual analytics approach combined with Statsitical Methods, demographic features which are traditionally related to Violent Crime Rate (VCR) are identified and OLS Univariate and Multivariate Regression are used as a precursor to GWR. VIF and pearson correlation statistics show strong colinearity in many of the traditionally used features and so human reasoning is used to rectify this. Bandwidth kernel smoothing size of 67 with a Bi-Square type is best for GWR. GWR and OLS regression shows that there is local variation in VCR and K-Means clustering using 5 cluster provides an effective way of seperating violent crime in London into 5 coherent groups.

---

## 1 PROBLEM STATEMENT

According to the BBC, in 2019, over 150 people were killed in London with 10% of these relating to domestic violence [1]. Similarly, 2020 represents a sixth consecutive year of London homicides rising above 100 [1].

While this coincides with 10 years of budget cuts to policing and community safety alongside wider budget cuts to communities [2], the government's behavioural insights team states that "violence in some boroughs is decreasing" and that increases in crime after a decade of decreases can be attributed in the increase in the use of knives [3]. The suggestion by this team is that violence in London occurs within small clusters within neighbourhoods with less than 3000 residents [3].

It has been argued that action taken in the past to reduce violent crime, such as the introduction of the 'knife ASBO', has not worked and serves to stigmatize and criminalise young vulnerable people in London [4]. To provide effective action, it is important to understand the causes of violent crime in London. As such, this study looks to answer the following questions:

- What is the distribution of violent crime in London and where are the highest rates of violent crime?
- Are there locally varying demographic factors that drive the distribution of violent crime in London in 2011?

The dataset used to answer these questions is the 2011 London Census which contains demographic data about the population of London. This also includes total crime numbers and the numbers of various crimes committed within London areas.

## 2 STATE OF THE ART

Investigations into local crime patterns inspired by ecological theories about crime have been done before using Geographically Weighted Regression (GWR). In 2007, Cahill and Mulligan explore the spatial patterns of crime with respect to demographic in Portland, Oregon using the Portland 2000 Census and averaged violent crime data collected from 1998 to 2002 [5]. Features from the 2000 Census are selected based on theories about the causes of violent crime and used to explore the drivers for violent crime in Portland. This is done first using a Global Ordinary Least Squares model, which assumes no local variation, and then a GWR for comparison. The OLS model provided support for criminal opportunity theory but left 60% of variance in violence unexplained which warranted the use of GWR. It was found that 4 out of 8 parameters displayed significant variation in space and using hierarchical clustering it was possible to assign areas in Portland to 6 coherent groups to explain how crime varies locally. They conclude that GWR can be particularly useful in policy studies allowing for different effective intervention in different areas. Using this methodology in the context of London could provide insights into how crime varies in London.

A study by Beecham et al., in 2018 on the locally varying explanations behind Brexit using UK census data takes a very similar approach. Using existing human theories about why people voted Leave, features were selected and global models, OLS models and GWR models were produced. They make more effective use of visual techniques than that of Cahill and Mulligan through the use of equal-area projections, scatter plots to explain feature choices and regression grids for each feature vs Leave vote to show how correlation varies between regions. However, it is important to note that Beecham et al., find that by smoothing over relatively large Local Authorities (LAs), local variation is lost, and the importance of social demographics is "likely overstated" [6]. This could be combatted in this study by using Area data rather than aggregating by LA.

Wang et al., 2019 proposed the use of GWR to estimate local crime rates in Chicago using demographic data coupled with large-scale Point-Of-Interest (POI) data and taxi flow data [7]. Nodal features such as POI and demographic features as well as Edge features which depict geographical influence are used for crime rate inference. The GWR framework is utilised with a negative binomial (NB) regression model (GWNBR) to guarantee that crime rate is a non-negative integer. Features such as poverty index, total population and ethnic diversity are produced using the 2000 US Census based on previous research on demographic features that cause crime. The Professional Index in this study is found to be fairly consistent with crime with a Pearson Correlation value of 0.3221 and a p-value of 0.0043. A Cross-validation approach is used in bandwidth parameter selection to estimate the best bandwidth for the data. Feature importance is shown primarily through choropleths with diverging colour scales. The GWNBR model consistently outperforms NB models but does not find strong correlations between violent crime and POI. This could be due to violent crime making up a small proportion of total crime.

## 3 PROPERTIES OF THE DATA

The dataset used contains 436 columns of demographic and crime data on 649 areas in London. The London Census was collected by the Office for National Statistics on the 27th March 2011 and is known to be "the most comprehensive data set on the population " [8].

It can be seen that while demographic data from the London Census is only available for 2011, crime data is available for financial years 2001-2012. In order to reduce the dimensionality of the data, features have been derived by grouping columns (such as age ranges) and then finding the percentage of the population that each group makes up. For example, the group 'children' contains the sum of the number of people aged between 0-14 and then has been divided by the total population for that area. For groups such as 'Professionals' we have calculated the percentage of people who are 'professionals' from the total number of people who are employed. Through this process of feature extraction, the dataset was reduced to have only 57 columns relating to the various demographic groups, total crime in London, percentage of crime that is violent crime (VCR), and population density for each area of London. All feature columns that are derived are real, numeric columns with columns for Area and Borough being Categorical.

Violent Crime Rate for each area in this study is defined in equation 1:

$$VCR = \frac{Total\ Number\ of\ Violent\ Crimes}{Total\ Number\ of\ Crimes} \quad (1)$$

Population density for each area is defined in equation 2:

$$Population\ Density = \frac{Total\ Population}{Area\ in\ Hectares} \quad (2)$$

It can be seen that the histogram for the 'commuting' feature (individuals travelling more than 10km for work) is bimodal. This could be as a result of having 2 distinct groups of areas with regards to commuting. Looking to understand this variation will drive the initial analysis. Observing the histograms, scatterplots and density plots for the features provided some initial investigation into the dataset. Figure 1 shows these plots for features of interest based on previous academic research [1] [2] [3] [4]. These were plotted for all features to decide on which features to use for analysis.

Boxplots and descriptive statistics for all columns show that there is an extremely large range in Total Crime in 2011 driven by high values of total crime in West End. The range of total crime in 2011 is 20636 with a mean of 1278. This results in extremely large variance in total crime in 2011. The reasons for this will be inspected during analysis. The range for area in hectares is 2867.4 with a mean of 251.5. This must be taken into account when any sort of geospatial smoothing is done as this provides the scale at which smoothing should be done to not overly smooth the data.

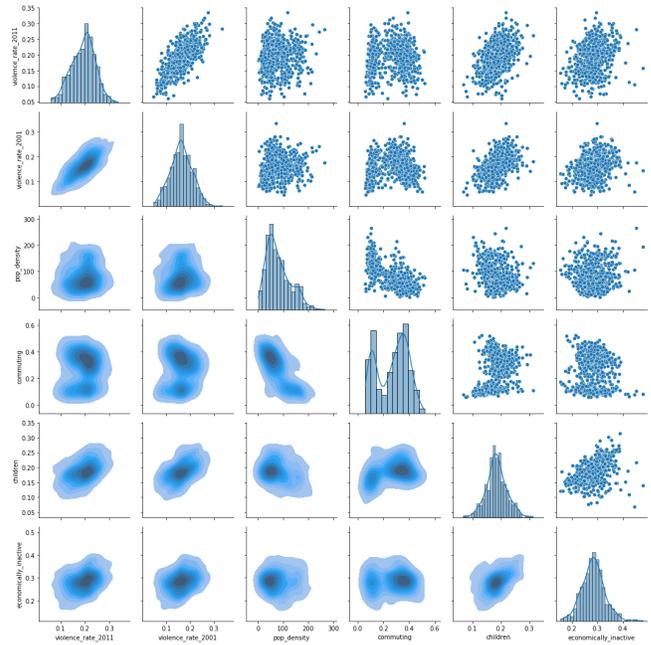

*Figure 1: Pairgrid plotted using Seaborn library showing variable histogram along diagonal, density plot on the lower portion of the grid and scatter plot on the upper portion of grid.*

A Shapefile containing the area boundaries for London is joined to the London Census data to allow for geospatial plotting of the data. These are joined by area name and dissolved by Borough for Local Authority scale investigation.

## 4 ANALYSIS

### 4.1 Approach

PRELIMINARY INVESTIGATION

After the initial aggregation and data transformation to percentage of population, data visualisation is used for investigation. This is done in two ways:
1. Choropleth maps looking at the spatial variation of features at Local Authority and Area level
2. Scatter plots and density plots to look at relationships between variables and with violence rate in 2011 (VCR)

In this process, human knowledge is used to pick out preliminary features for analysis.

Following this, Multidimensional Scaling (MDS) is used to check the quality of feature choices. The selected features are projected onto 2 axes with the colour scale set as the VCR. The distance between points indicates how similar the areas are and human reasoning is used to assess whether there is consistency between the projection plot and VCR [9]. Should according to the visualisations produced.

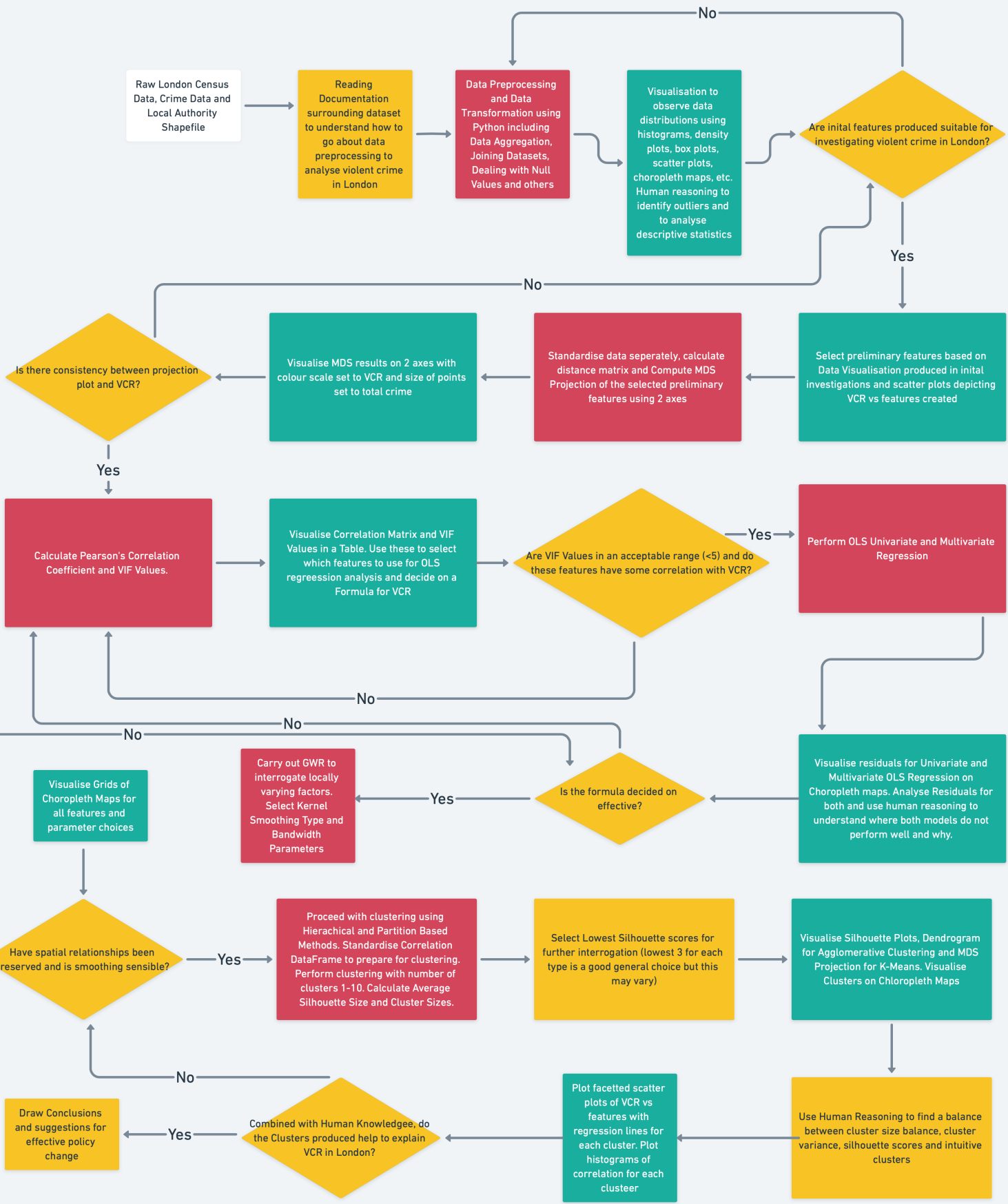

*Diagram 1: Visual Analysis Workflow. Red boxes are tasks which require computational processing, Teal boxes are data visualization tasks, Yellow Boxes/Diamonds are tasks assigned to the human analyst.*

## CORRELATION MATRICES, VARIANCE INFLATION FACTORS AND ORDINARY LEAST SQUARES (OLS) REGRESSION

A global OLS regression is used as a comparison and a precursor to GWR [5] [6]. The correlation matrix between features helps to further refine the features selected for analysis.

Variance Inflation Factors (VIF) are used to assess multi-colinearity within features [10]. In this study, VIFs greater than five are regarded as having severe multi-collinearity [10]. If there are 2 or more variables which show high VIF, human knowledge is used to decide which to remove from analysis. Multivariate and Univariate regressions are evaluated by plotting residuals for each area on choropleth maps. Human perception is used to decide where the models overpredict/underpredict. Refinement of feature choices is iterated over and a formula for VCR is inferred.

## GEOGRAPHICALLY WEIGHTED REGRESSION

GWR examines the spatial variation in the correlation of VCR to the features. In this process, parameter choices in bandwidth and kernel smoothing type are essential. GWR is done with different parameter combinations to ensure strong parameter choices. Choropleth maps after GWR are plotted to see the effects that parameter choices have. Human reasoning is used to decide on parameter choices based on how these maps reveal spatial patterns.

## HIERACHICAL/PARTITION-BASED CLUSTERING

Correlation coefficients for each feature in each area is split into subsets using clustering. Partition-based and Hierarchical clustering are both appropriate for this. K-Means and Agglomerative Clustering are both tested on.

Silhouette scores and cluster maps are used to decide on the number of clusters. For Agglomerative clustering, a dendrogram is used to understand how the clusters vary. Human reasoning is used when observing choropleth maps to see which number of clusters and which algorithm yields the most insightful results. Therefore, number of clusters and technique used is a balance between statistical reasoning (minimizing silhouette scores) and human understanding (clusters that intuitively make sense). If there is no number of clusters that make sense, the features used in GWR is revisited.

To draw conclusions about locally varying factors that influence VCR in London, the features are plotted against VCR and faceted by cluster. Regression is done for each cluster to view how these vary. Histograms of correlation with each feature within each cluster is also plotted. Human knowledge is required to draw conclusions based upon these visualisations.

### 4.2 Process

## PRELIMINARY INVESTIGATION

Figure 2 shows choropleth maps of a selection of the most significant features. The plot of standard deviation in VCR shows how the VCR varies within LAs. The anomalously high standard deviation in Westminster coincides with extremely high total crime numbers in the borough. This high standard deviation seems to be driven by large numbers of crime in tourist hotspots such as West End where other crimes (such as theft or fraud) dominate total crime numbers which causes VCR to be lower. When the region of Westminster is plotted in isolation, it can be seen that North-Western areas (Queen's Park, Harrow Road, Westbourne, Church Street) have higher VCR which has been explained by violent burglaries which take place in these affluent areas [11]. The high standard deviation in VCR of Westminster can therefore be explained by large volumes of theft and other crimes in tourist areas coupled with higher numbers of violent burglary in affluent residential areas. There is also very clear spatial relationship between population density and commuting. Densely populated areas in central London have low proportions of commuters. This explains the bimodal nature of the 'Commuting' feature. During feature selection, it should be noted that these two variables are closely linked.

Following this line of interrogation, scatter plots produced for features vs VCR were coloured by %Commuting. This is to see whether areas within central London, where commuting is less common, held different VCR relationships with the features. The areas are sized by total crime to examine the relationship with total crime in each area. Observing Figure 3, there are fairly clear relationships between VCR and a number of the features.

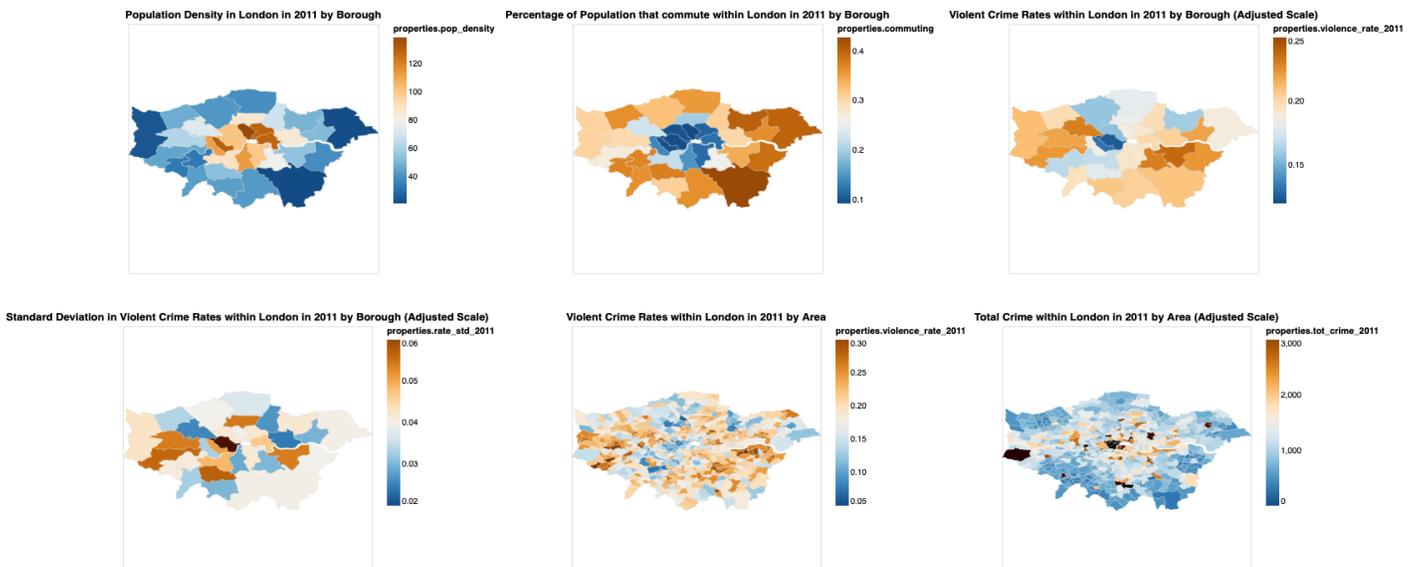

*Figure 2: Choropleth's depicting various features of interest created using Altair and Geopandas*

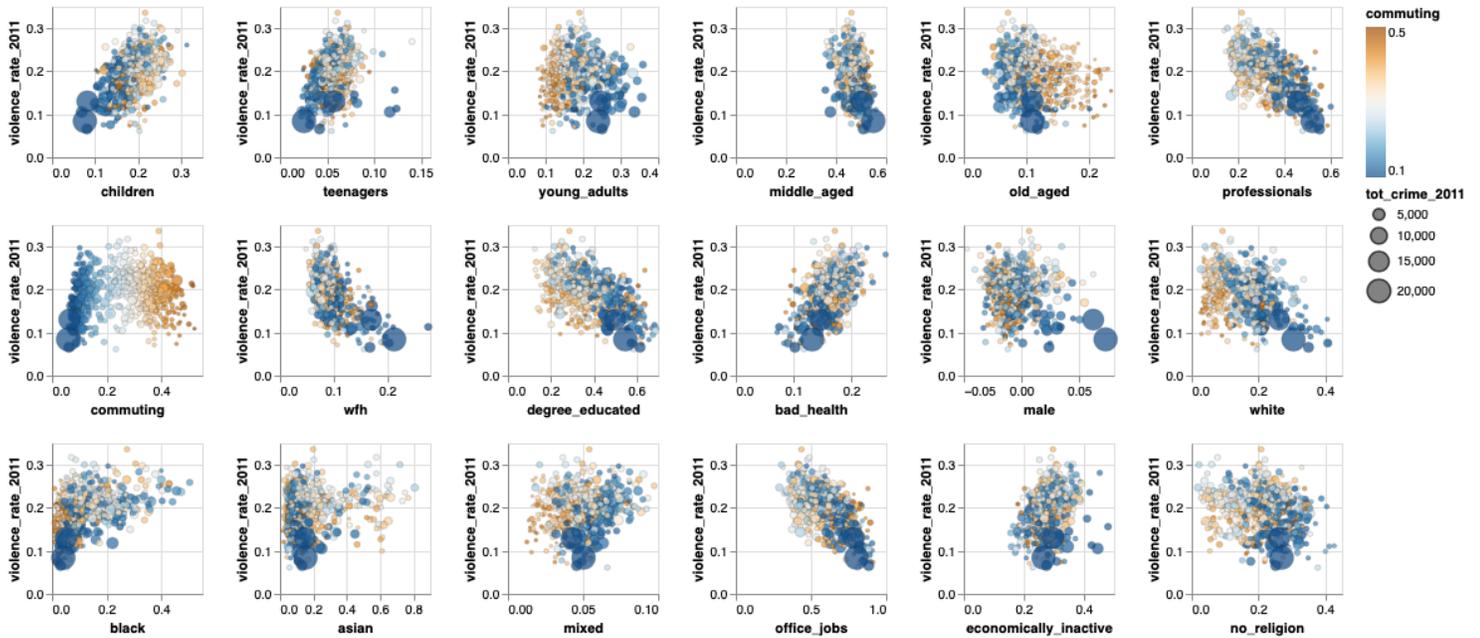

*Figure 3: Scatter plot grid of features vs VCR, Coloured by %Commuting and sized by total crime.*

In sociological literature, it is noted that the root causes of violent crime are poverty, inequality and financial deficiency [12]. Using this, the following features were chosen for further investigation based on human knowledge and perceived numerical relationship: %Children, %Teenagers, %Degree Educated, %WFH, %Bad Health, %White, %Office Jobs, %Professionals, %Economically Inactive, Population Density. Within the 'White' feature, it can be seen that in non-commuter areas, the feature has a stronger relationship with VCR. In the 'Old Age' feature a similar relationship can be seen but this hasn't been selected due to less strong perceived correlation with VCR.

Figure 4 shows these features projected onto two axes interactively using MDS showing there is a clear spatial relationship with VCR. Promisingly, there are no real outliers on the projection. At the top left-hand side of the projection, there are areas with higher VCR. The areas here are areas outside of central London such as Southall Green, Stonebridge and East India and Lansbury, In the bottom-right, where VCR is low, there are central areas such as Courtfield, Pembridge and Muswell Hill. This is a key indication that the features selected can be used to investigate VCR in London. Discovering what makes areas with high VCR similar to each other is the next stage of this analysis.

## CORRELATION MATRICES, VARIANCE INFLATION FACTORS AND ORDINARY LEAST SQUARES (OLS) REGRESSION

Using a correlation matrix, it is possible to see colour coded Pearson correlation coefficients for all of the preliminary features. Doing so revealed that there is strong multi-colinearity in a number of features. Namely, the feature '%Professionals' had a 0.94 correlation with '%Degree Educated', 0.96 with 'Office Jobs' and 0.76 with 'WFH'. Initially, both '%Degree Educated' and '%Professionals' were used as these were to portray different aspects of the population (%Degree Educated is likely to include newly graduated students whereas %Professionals represents the older middle classes). However, when both were used, this resulted in VIF scores of 21.7 and 17.0 respectively. As such, only the '%Professionals' feature was used. Similarly, the '%Children' metric was used instead of the '%Teenagers' as a proxy for the number of families there are in each area. It has been claimed that violent crime is being perpetrated by increasingly young individuals [3]. This feature would allow this to be investigated. Selected features selected are:

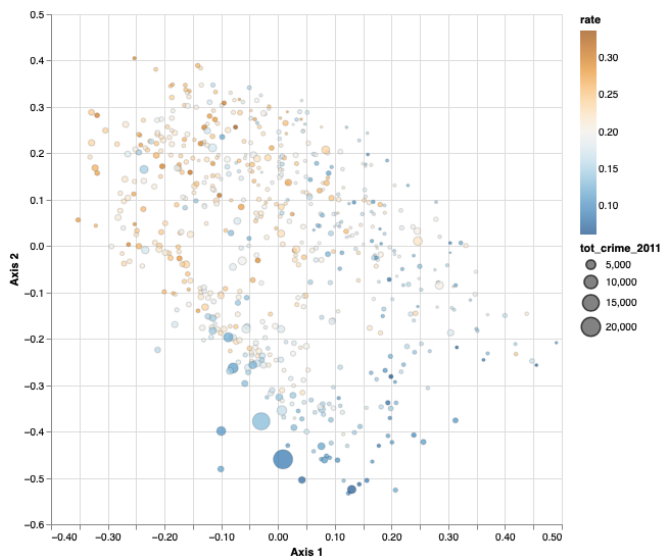

*Figure 4: 2 Axes MDS Projection of preliminarily selected features plotted interactively using Altair. Colored by VCR and sized by total crime*

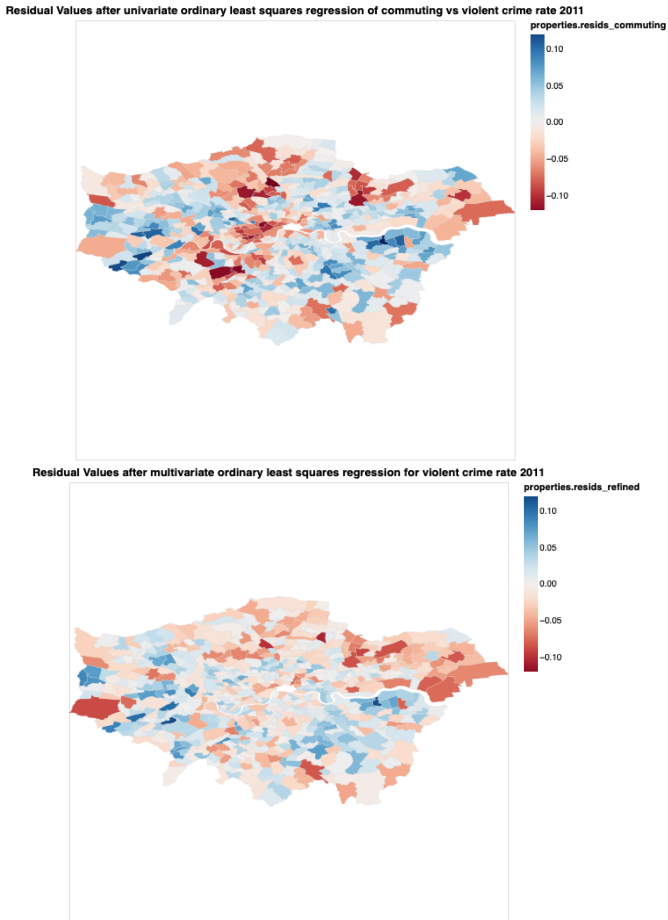

Figure 5: Choropleths of residuals for OLS Regression modelling plotted using fixed diverging color scale

- %White: VIF = 1.98
- %Economically Inactive: VIF = 3.74
- %Bad Health: VIF = 3.20
- %Commuting: VIF = 1.66
- %Professional: VIF = 1.83
- %Children: VIF = 2.06

Analysis using scatter plots, choropleth maps, correlation matrix and descriptive statistics combined with specialist knowledge suggests the following formula for VCR in London.

$$VCR \sim \%Children + \%Bad\ Health + \%Professionals + \%White + \%Economically\ Inactive + \%Commuting$$

Residual maps for both univariate OLS regression and multivariate (using the above formula) OLS regression show that using only '%Commuting', VCR in Central London is generally over predicted with some areas of North-East London being overpredicted also. Using a multivariate regression shows a different story where suburban areas in East, North and West London are generally overpredicted, but less so. The lighter shades across London in the multivariate OLS indicates that there are a number of factors that correlate to VCR. The multivariate regression strongly overpredicts VCR in Valley (Waltham Forest) and Alexandra (Harringey). This indicates that these factors could be locally varying and there is a case for GWR.

### GEOGRAPHICALLY WEIGHTED REGRESSION

The results from the regression modelling, scatter plots and inspection of features on choropleth map all point towards locally varying factors that influence VCR. This justifies the use of GWR in this study. When performing GWR, kernel bandwidth and smoother type makes a big difference to the results. The smoother size had to be appropriate for the scale of the spatial patterns that is being investigated. Kernel smoother types were all tested (Gaussian, Bi-square and Exponential). Kernel smoother bandwidth was investigated from size 1-200 in increments of 25 before deciding on an area of 50-75 to investigate further. Fixed smoothing was decided upon as this theoretically makes the most sense when looking at spatial relationships rather than density-based relationships. Adaptive smoothing uses number of data points rather than a fixed area.

The results from testing can be seen in Figure 5. When observing these maps, it is clear to see that the bi-square kernel smoother provides a much smaller and sharper smoothing area due to how the smoother is shaped. Using larger bandwidth values loses the variance between areas (as seen by the change in scale) as a result of too much smoothing. Smoothing with small bandwidth using a bisquare smoother results in barely any smoothing taking place which is seen by no noticeable patterns in correlation to VCR. Viable candidates for final smoothing parameters are therefore small bandwidth Gaussian smoother or medium bandwidth bisquare smoother. The medium bandwidth (67) bisquare smoother preserves some of the higher correlation to crime rate in North-West London and this is why this combination was chosen for final GWR.

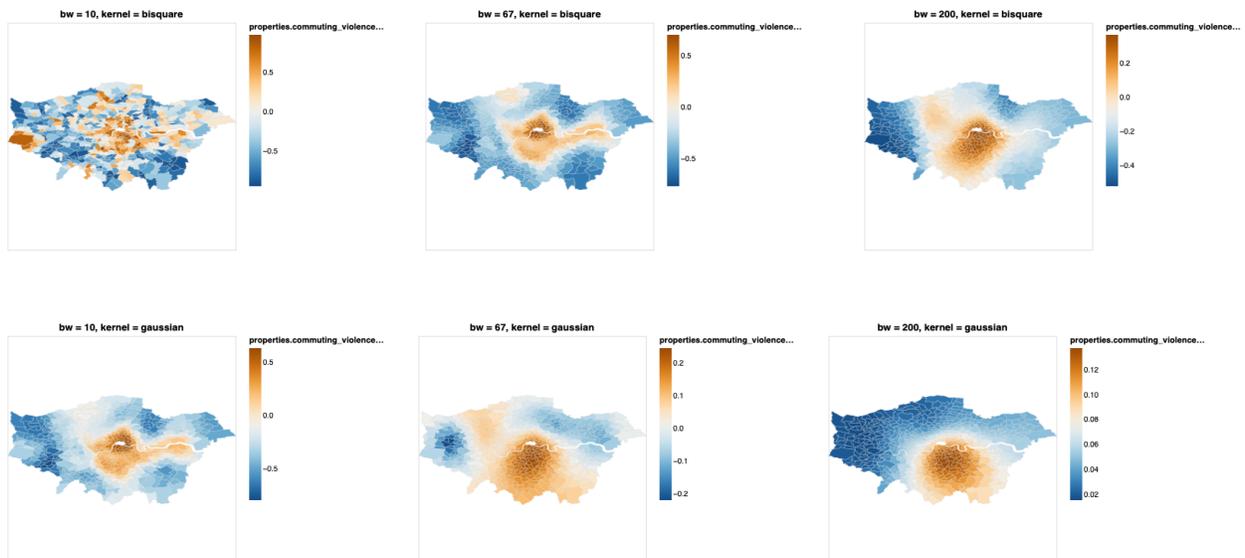

Figure 6: Results of GWR Parameter testing plotted for %Commuting feature

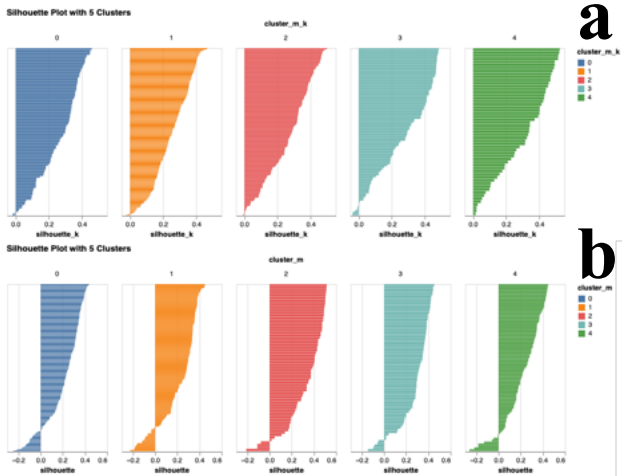

Figure 7: Silhouette plots for each cluster calculated using SKLearn's Silhouette_score function. a) for K-Means b) for Agglomerative Clustering

Studying the maps for all features, it could be seen that the spatial patterns for %Bad Health and %Degree Educated were the inverse of one another. As these are both features which are a proxy for affluence, it could be suggested that these are areas where the potential key driver for VCR is income inequality. To prevent giving undue weight to affluence in the GWR, only %Degree Educated will be used in clustering. %White has a larger correlation to crime rates in East and South-East London and a lower correlation in Central London Areas. These areas also have moderate correlation between %Children and VCR. This indicates that these are areas filled with working class, diverse families. These are areas where knife crime and gang violence are prevalent and so these may be drivers for VCR. %Economically inactive shows higher correlation in North-East London and no correlation in Central London. It is hard to predict why this may be the case but one potential cause in these areas may be poverty. Therefore, the final features chosen for clustering are: %White, %Commuting, %Degree-Educated, %Children, %Economically Inactive.

### HIERACHICAL/PARTITION-BASED CLUSTERING

When testing K-means and Agglomerative clustering, silhouette scores as well as human knowledge in understanding the resulting clusters was used. After testing both algorithms with 1-10 clusters, it was found that the best clustering was found with 5 or 6 clusters for both algorithms.

A dendrogram (Figure 8) was plotted to understand the difference between clusters for the Agglomerative clustering and it was found that there was little difference between 4 and 5 clusters but moderate vertical difference between 5 and 6. However, the dendrogram showed that the results were dominated by one large group with 275 areas within it.

In general, the K-means algorithm provided higher silhouette scores than the Agglomerative algorithm but had a much higher proportion of positive values as shown in Figure 7. When plotting clustering results on a choropleth map, the 5 cluster K-Means made the most intuitive sense as it was able to define the Suburban areas (Cluster 1), commuting towns (Cluster 0), high-inequality commuting areas (Cluster 2), tourism centre of London (Cluster 4) and transition areas (Cluster 3) well. This version had a slightly higher silhouette score and worse balance across cluster sizes than the 6 cluster K-means but due to much better intuitive clusters, the 5 cluster K-Means was preferred. Projecting the 5 cluster K-Means onto 2 axes using MDS showed that clusters 0 and 1 were the most similar with others being roughly evenly distanced. This is interesting as cluster 0 could be classed as commuter towns of London whereas cluster 1 looks to be suburban areas. Conceptually, it is easy to see how these may

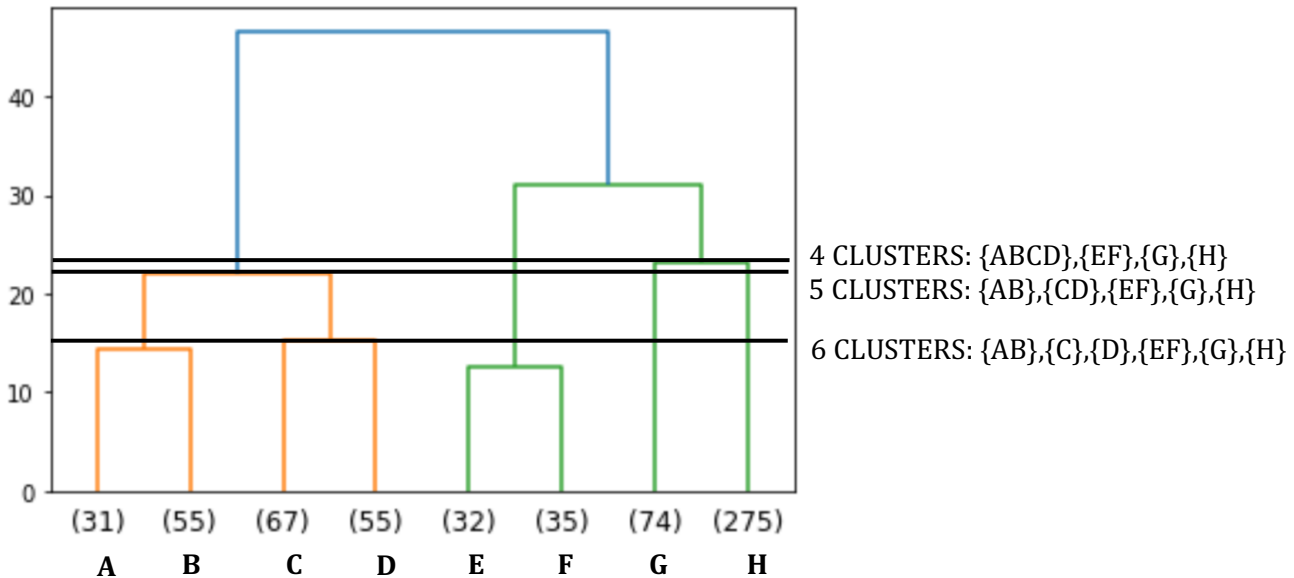

Figure 8: Dendrogram showing similarity and size of clusters produced by agglomerative clustering

be similar

### 4.3 Results

VCR in London in 2011 can be grouped into 5 distinct groups (Figure 9a) which vary spatially according to demographics of areas of London. K-Means clustering performs better than Agglomerative clustering (Figure 9b). The area with the highest VCR is Glyndon (Greenwich) with the Greenwich Borough having the highest VCR. %Commuting alone is unable to explain VCR in London.

VCR in London varies locally, and Figure 10 shows VCR regressed against features faceted by cluster. Here it is possible to see how VCR varies according to demographic in different subsets of areas in London. In London's Tourism Centre, VCR has a strong positive relationship with proportion of people that commute unlike other areas. Similarly, in more affluent residential areas of London (Cluster 0), %Children has no relationship with VCR but strong negative correlation with %Economically inactive.

Using these results, it is possible to target violent crime in areas of London differently as a means to more effective action. For example, in areas where there is a strong correlation between %Children and VCR, it is possible that crimes are being committed by younger individuals as a result of family income inequalities. Investment in youth centres may be effective in these areas.

### 5 CRITICAL REFLECTION

While it was possible to examine the distribution of VCR in London and understand where violent crime happens in London, this style of analysis makes it difficult to identify the root causes of violent crime in London. In the results, it was suggested that areas with high correlation between number of children and VCR could have high numbers of young offenders and investment in youth centres and community centres could decrease violent crime rates. As a limitation of using only VCR in London, it is hard to assign cause to high correlation. High correlation between children and violent crime could be caused by a number of external factors. Therefore, while it is possible to look at locally varying factors that relate to VCR, it is hard to explain the characteristics that drive violent crime.

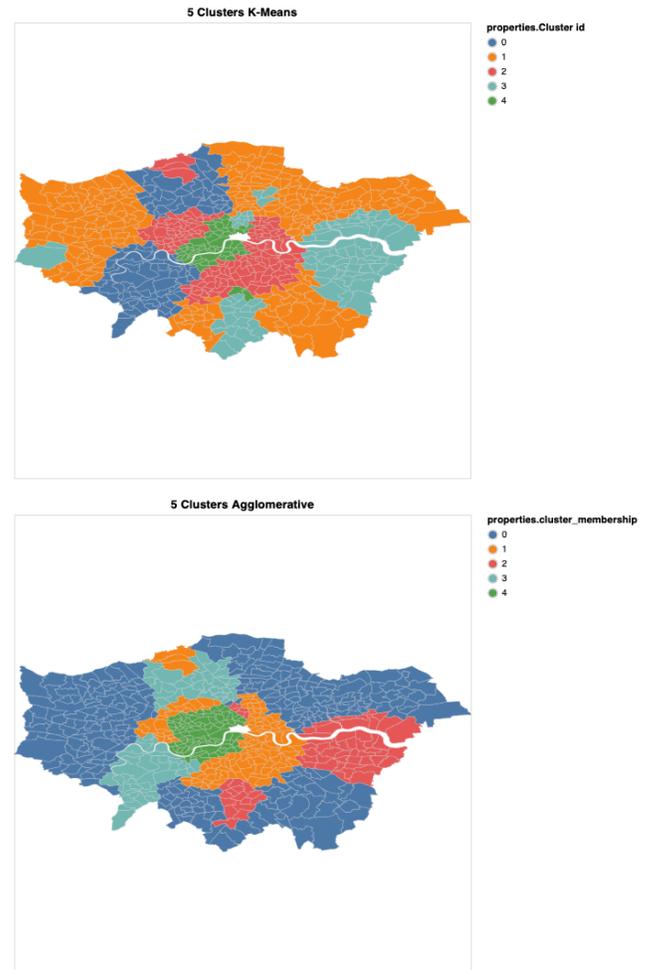

*Figure 9: Clustering results for K-Means and Agglomerative algortihms with 5 Clusters. Cluster assignment described in previous section*

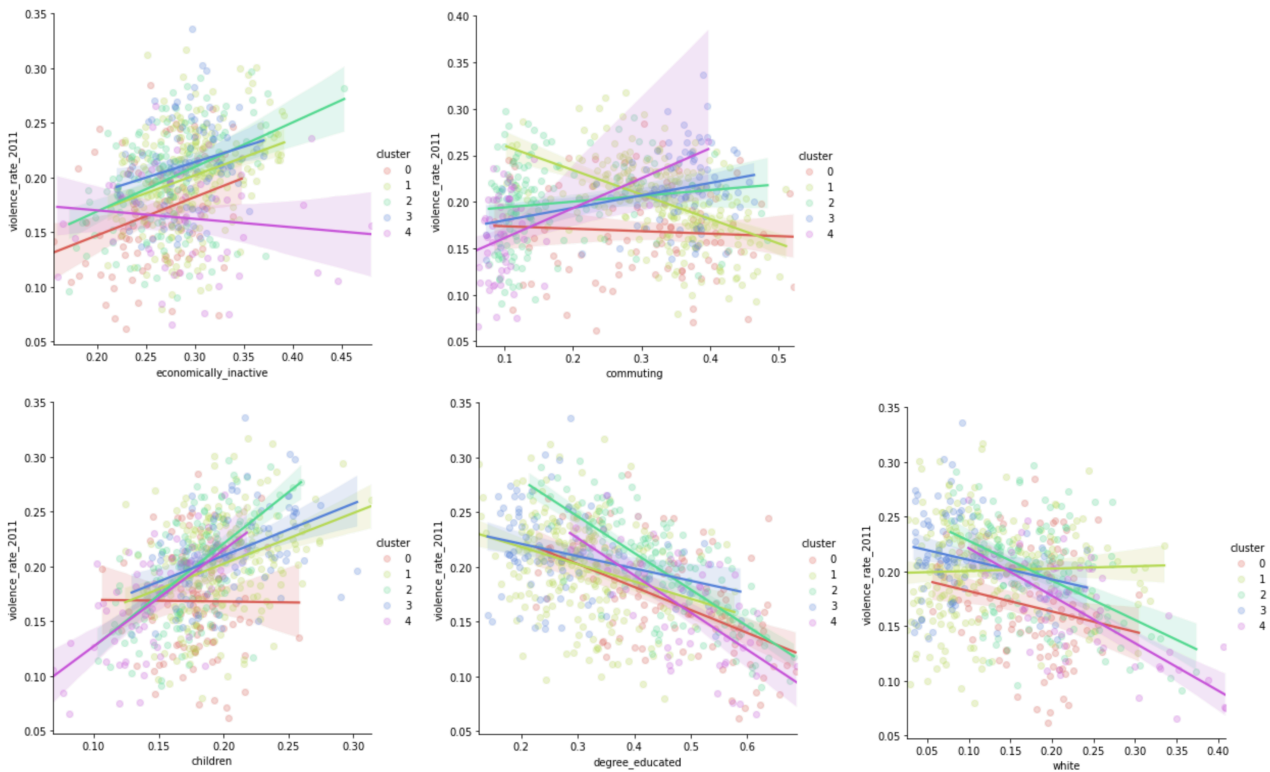

*Figure 10: Features vs VCR with regressions faceted by cluster plotted using Seaborn's lmplot function.*

One pitfall of the analysis conducted is that it assumes linear relationships between the variables where this may not always be the case. For example, the 'WFH' variables looks as though it could share a 2$^{nd}$ order polynomial relationship with VCR. The iterative approach to analysis combined with statistical methods ensured that colinearity in my analysis was minimised and that appropriate features for analysis were selected. This is an important use of visual analytics as without appropriate variables, it would be easy to find misleading results and give undue weight to particular characteristics of the population. This study focuses on crime at an area level rather than an individual level and only applies to violent crime in London so cannot easily be generalised to other urban capitals.

Appropriate use of colour scales and visual representation helped to pick out important relationships between variables. An example of where this was done is when plotting MDS projections to check whether there was a relationship between similarity between features and VCR. Fixed colour scales allowed comparison between the residual maps for OLS regression.

To further this analysis, one could observe how violent crime rates have changed in time, in line with budget cuts and the 2008 financial crisis. The DBSCAN algorithm for density-based clustering could be used to pick out areas that have similar changes of VCR in time. Doing this, it would be possible to look at how different types of areas have been affected by budget cuts to policing and communities.

In addition, while non-stationarity of the spatial relationships between the features and VCR has been investigated, the scale in which these processes may take place can vary with scale. It is possible to do Multi-Scale GWR which "derives an optimal bandwidth vector" [13] to find the scale at which processes can take place. This would help investigate these features while acknowledging that these factors may locally vary with VCR at different scales.

Lastly, instead of using Agglomerative and K-Means Clustering which are Hard Clustering methods, Fuzzy Clustering techniques could be used to acknowledge that cluster assignment is fluid. This means that areas can belong to more than one cluster. This would require more robust outlier removal as fuzzy clustering techniques are sensitive to noise and outliers but could provide insight into how clustering varies spatially [14].

**Table of word counts**

| Problem statement/ | 251 |
|---|---|
| State of the art | 525 |
| Properties of the data | 499 + 1 Figure |
| Analysis: Approach | 524 + 1 Diagram |
| Analysis: Process | 1549 + 7 Figures |
| Analysis: Results | 199 + 2 Figures |
| Critical reflection | 519 |